\documentclass{jpconf}
\usepackage{cite}
\usepackage{graphics}

\begin{document}

\title{Magnetic interactions and high-field properties\\ of Ag$_2$VOP$_2$O$_7$: frustrated alternating chain\\ close to the dimer limit}

\author{Alexander A Tsirlin$^{1,2}$, Ramesh Nath$^{1}$, Franziska Weickert$^{1,3}$,\\ Yurii Skourski$^{3}$, Christoph Geibel$^{1}$, and Helge Rosner$^{1}$}
\address{$^1$ Max Planck Institute for Chemical Physics of Solids, 01187 Dresden, Germany}
\address{$^2$ Department of Chemistry, Moscow State University, 119992 Moscow, Russia}
\address{$^3$ Dresden High Magnetic Field Laboratory, 01328 Dresden, Germany}
\ead{altsirlin@gmail.com, Helge.Rosner@cpfs.mpg.de}

\begin{abstract}
We report on high-field magnetic properties of the silver vanadium phosphate Ag$_2$VOP$_2$O$_7$. This compound has a layered crystal structure, but the specific topology of the V--P--O framework gives rise to a one-dimensional spin system, a frustrated alternating chain. Low-field magnetization measurements and band structure calculations show that Ag$_2$VOP$_2$O$_7$ is close to the dimer limit with the largest nearest-neighbor interaction of about 30 K. High-field magnetization data reveal the critical fields $\mu_0H_{c1}\simeq 23$ T (closing of the spin gap) and $\mu_0H_{c2}\simeq 30$ T (saturation by full alignment of the magnetic moments). From $H_{c1}$ to $H_{c2}$ the magnetization increases sharply similar to the system of isolated dimers. Thus, the magnetic frustration in Ag$_2$VOP$_2$O$_7$ bears little influence on the high-field properties of this compound.
\end{abstract}

Low-dimensional and, possibly, frustrated spin systems present numerous quantum phenomena dealing with both the ground state and finite-temperature properties. An external magnetic field alters the physics of such systems and induces a number of additional features. The field causes spin excitations which are naturally treated as bosons. These bosons can undergo Bose-Einstein condensation or form ordered structures (Mott-insulating phases)~\cite{BEC-review}. The latter scenario gives rise to plateaus in the magnetization curves. The observation and interpretation of such plateaus remains one of the attractive problems related to quantum phase transitions in low-dimensional spin systems~\cite{srcu2(bo3)2,azurite-2005,azurite-2008}.

\begin{figure}
\includegraphics{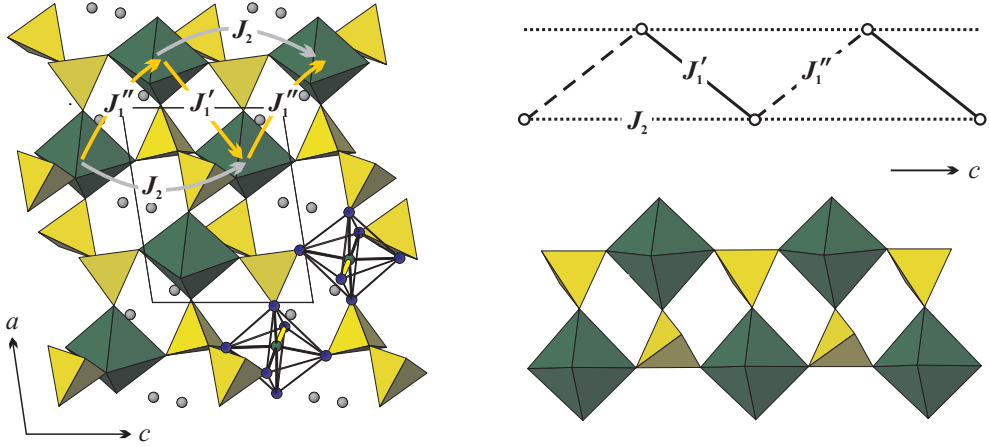}
\caption{\label{fig-structure}
Crystal structure of Ag$_2$VOP$_2$O$_7$ (left panel); the frustrated alternating chain model and the respective part of the structure (right panel). In the left panel, short V--O bonds are shown as thick yellow lines in the two VO$_6$ octahedra with removed faces. Solid, dashed, and dotted lines in the right panel denote $J_1'$, $J_1''$, and $J_2$, respectively.}
\end{figure}

The formation of bosonic Mott-insulating phases (and the respective magnetization plateaus) has been predicted for a wide range of spin models including Heisenberg models for different lattice types. One of the latter models is a frustrated alternating chain, i.e., a chain with the alternating nearest-neighbour interactions $J_1',J_1''$, and the next-nearest-neighbour interaction $J_2$ (see the right panel of Figure~\ref{fig-structure})~\cite{totsuka,haga}. Despite considerable theoretical effort in studying this model (see Ref.~\cite{totsuka} and references therein), experimental results are scarce due to the lack of appropriate model compounds. To the best of our knowledge, only two materials have been suggested as frustrated alternating chain systems: the low-temperature, spin-Peierls phase of CuGeO$_3$~\cite{cugeo3-1995} and Cu$_2$(C$_5$H$_{12}$N$_2$)Cl$_4$ (copper 1,4-diazacycloheptane chloride, abbreviated as CuHpCl)~\cite{cuhpcl-1990,cuhpcl-1997}. However, both the compounds do not show the desired high-field physics of the Heisenberg model for the frustrated alternating chain. In case of CuGeO$_3$, spin-lattice coupling is essential and produces the high-field soliton phase with incommensurate modulation of the lattice~\cite{cugeo3}. As for CuHpCl, one also has to extend the simple Heisenberg model and to consider exchange anisotropy in order to interpret the experimental results~\cite{cuhpcl-2006}. Thus, there are no appropriate model compounds, and the search for new systems is desirable.

Recently, we have proposed silver vanadium phosphate, Ag$_2$VOP$_2$O$_7$, as a frustrated alternating chain system~\cite{ag2vop2o7-2008}. In the following, we briefly review the miscroscopic model of exchange interactions in this compound and present new high-field magnetization data. 

The crystal structure of Ag$_2$VOP$_2$O$_7$~\cite{ag2vop2o7} is visualized in the left panel of Figure~\ref{fig-structure}. Vanadium atoms are located in distorted VO$_6$ octahedra with short vanadyl bonds. The octahedra are joined into layers via PO$_4$ tetrahedra, and silver cations are located between the layers. Vanadium atoms have the oxidation state of +4 implying one unpaired $d$ electron per magnetic atom. The distortion of the octahedral coordination forces the single $d$ electron to occupy the $d_{xy}$ orbital lying perpendicular to the short bond and to form a non-degenerate orbital ground state. The specific orbital state of vanadium limits the number of possible superexchange pathways and gives rise to a one-dimensional spin system~\cite{ag2vop2o7-2008}. The respective empirical model is a frustrated alternating chain (see the right panel of Figure~\ref{fig-structure}).

Single-phase polycrystalline samples of Ag$_2$VOP$_2$O$_7$ were prepared by solid-state technique as described in Ref.~\cite{ag2vop2o7-2008}. The magnetization measurements were carried out at the Dresden High Magnetic Field Laboratory using a pulsed magnet energized by a 50 MJ, 24 kV capacitor bank. The characteristic duration of the pulse was 50 ms for magnetic fields up to 60 T. Low temperature of 1.5 K was obtained by a $^{4}$He-flow cryostat installed in the bore of the magnet coil. The high precision magnetometer consists of a compensated pick-up coil system within a teflon tube of 2.5 mm in diameter where the sample was mounted. The signal of the pick-up coil system was measured twice during the magnet pulse, with and without the sample. The integrated difference is proportional to the sample magnetization.

In Ref.~\cite{ag2vop2o7-2008}, we employed low-field magnetization measurements and band structure calculations to confirm the empirical model and to estimate individual exchange couplings. We have found that the system is gapped, $J_1'$ is $30-35$ K, while $J_1''$ and $J_2$ do not exceed 5 K. Interchain couplings are even weaker, below 1 K. Using the representative values $J_1'=35$ K and $J_1''=J_2=5$~K, one finds the frustration ratio $\alpha=2J_2/(J_1'+J_1'')\simeq 0.25$ and the dimerization ratio $\delta=(J_1'-J_1'')/(J_1'+J_1'')\simeq 0.75$. The $\alpha$ and $\delta$ values can be directly mapped on the theoretical phase diagrams in Refs.~\cite{totsuka,haga} yielding the possible bosonic Mott-insulating phase (and the respective magnetization plateau) in Ag$_2$VOP$_2$O$_7$.

\begin{figure}
\begin{minipage}{9.2cm}
\includegraphics{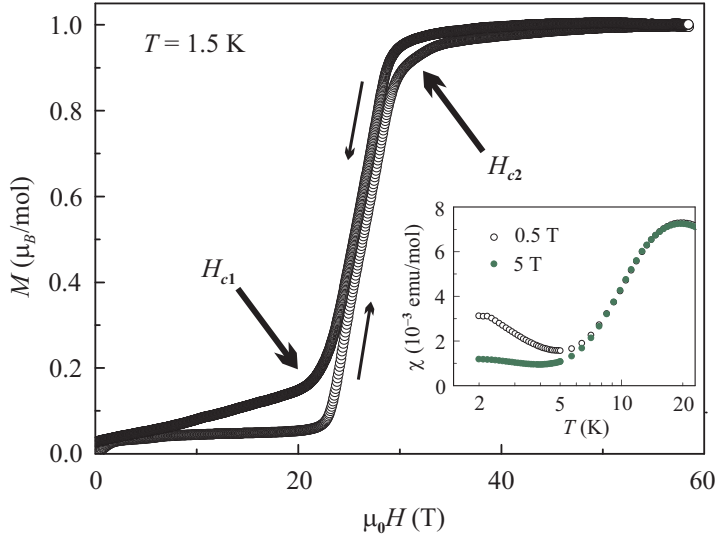}
\end{minipage}
\hspace{0.6cm}
\begin{minipage}[c]{6cm}
\caption{\label{fig-mvsh}
High-field magnetization of Ag$_2$VOP$_2$O$_7$ measured at 1.5~K. Thick arrows point to the phase transitions at $H_{c1}$ (closing of the spin gap) and $H_{c2}$ (saturation by full alignment of the magnetic moments), while thin arrows denote the data collected on the increase and the decrease of the field. The inset shows temperature-dependent magnetic susceptibility ($\chi$) measured in the applied fields of 0.5 and 5 T.}
\end{minipage}
\end{figure}

The high-field magnetization curves are shown in Figure~\ref{fig-mvsh}. The data collected on the increase and the decrease of the field reveal a notable hysteresis which is likely caused by a magnetocaloric effect. This effect emerges at the sharp increase of the magnetization between $H_{c1}$ and $H_{c2}$ giving rise to the heating of the sample and the smearing of the bends at $H_{c2}$ and $H_{c1}$ for the curves measured on the increase and the decrease of the field, respectively. In the following analysis, we rely on the former curve at low fields ($H\leq H_{c1}$) and on the latter curve at high fields ($H\geq H_{c2}$).

At low fields, the magnetization increases, shows a bend at about 2 T and stays constant ($\sim 0.04\ \mu_B$/mol) up to $\mu_0H_{c1}\simeq 23$ T. The bend at 2 T is likely caused by the saturation of paramagnetic impurities. This effect is also observed as the suppression of the low-temperature paramagnetic upturn in the susceptibility data at 5 T (see the inset of Figure~\ref{fig-mvsh}). The low magnetization up to 23 T is an indication of the spin gap in Ag$_2$VOP$_2$O$_7$. The gap is closed at $\mu_0H_{c1}\simeq 23$ T, and above $H_{c1}$ the magnetization increases sharply up to $\mu_0H_{c2}\simeq 30$ T. At $H_{c2}$, the moments are aligned ferromagnetically, and the magnetization remains nearly constant with further increase of the field. The shape of the magnetization curve is similar to that of other strongly dimerized spin systems. For isolated dimers, $H_{c1}$ and $H_{c2}$ coincide, since the spin gap is equal to the energy of the intradimer exchange interaction~\cite{bonner}. Additional interdimer interactions cause the difference between $H_{c1}$ and $H_{c2}$ (see, e.g., \cite{ba3cr2o8}). Yet the linear increase of the magnetization in this field range indicates the absence of the bosonic Mott-insulating phase and of the magnetization plateau.

The $H_{c1}$ value provides a direct estimate of the spin gap in Ag$_2$VOP$_2$O$_7$: $\Delta=g\mu_B\mu_0H_{c1}/k_B\simeq 31$ K, where we use the gyromagnetic ratio $g=2$. Unfortunately, we can not perform an exact comparison with the values expected from our set of the exchange coupling constants, since there is no general relation for $\Delta$ as a function of $\alpha$ and $\delta$. However, we can make a quantitative comparison by simplifying our spin model and neglecting $J_2$. For a non-frustrated $J_1'-J_1''$ alternating chain with $J_1'=35$ K, $J_1''=5$ K, the spin gap is about 32 K~\cite{johnston} in remarkable agreement with the experimental value of $\simeq 31$ K. Thus in Ag$_2$VOP$_2$O$_7$, $\Delta$ is nearly equal to the strongest coupling $J_1'$ consistent with the strong dimerization regime $J_1'\gg J_1'',J_2$. 

The saturation field $H_{c2}$ is a characteristic of the energy scale of the exchange couplings in Ag$_2$VOP$_2$O$_7$. There are no theoretical results available for the saturation field of the frustrated alternating chain, but we can make a rough comparison using simple, classical treatment of the model. The saturation implies the transition to a fully polarized, ferromagnetic state; therefore, $H_{c2}$ corresponds to the energy difference between the ground state and the ferromagnetic state. Band structure calculations~\cite{ag2vop2o7-2008} reveal that the ground state is composed of antiferromagnetic dimers (formed by $J_1'$), and the dimers are coupled antiferromagnetically by $J_2$. Within the classical Heisenberg model, the energy difference is $J_1'-J_1''+2J_2$, and our set of exchange couplings results in $\mu_0H_{c2}\simeq 29.8$ T in perfect agreement with the experimental value of 30 T. 

Thus, our set of the exchange coupling constants for Ag$_2$VOP$_2$O$_7$, as derived from low-field magnetization measurements and band structure calculations~\cite{ag2vop2o7-2008}, is consistent with the high-field magnetization data. This set predicts the formation of the bosonic Mott-insulating phase and the respective magnetization plateau at half-saturation. However, we do not observe the plateau in the high-field magnetization curve. One may speculate on several reasons for this discrepancy. First, our estimates of $J$'s are not exact, and a slight change of $\alpha$ and $\delta$ may shift the system out of the desired plateau region. Second, the theoretical results are not exact as well: Refs.~\cite{totsuka,haga} suggest slightly different boundaries of the plateau region. Finally, non-zero interchain couplings may play a certain role in stabilizing/destabilizing the high-field phase. Probably, one should search for a frustrated alternating chain compound with weaker dimerization in order to safely reach the Mott-insulating phase of this model and to observe the respective magnetization plateau.

In summary, we studied high-field magnetic properties of the frustrated alternating chain compound Ag$_2$VOP$_2$O$_7$. The magnetization curve reveals a spin gap of about 31 K and a saturation field of about 30 T. These values are in perfect agreement with our previous estimates of the exchange couplings in Ag$_2$VOP$_2$O$_7$~\cite{ag2vop2o7-2008}. As the spin gap is closed, the magnetization increases in a linear fashion up to the saturation. No features related to the possible bosonic Mott-insulating phase were observed. 

\ack
The authors acknowledge the financial support of GIF (Grant No. I-811-257.14/03), RFBR (Project No. 07-03-00890), and the Emmy-Noether-Program of the DFG. A.Ts. is grateful to MPI CPfS for hospitality and funding during the stay.

\section*{References}

\providecommand{\newblock}{}

\end{document}